\documentclass[final,1p,times,twocolumn,authoryear]{elsarticle}

\usepackage{url}
\usepackage{graphicx}
\usepackage{subfigure}
\usepackage{amsmath, amsthm, amssymb, amsfonts}
\usepackage{ragged2e}
\usepackage{rotating}

\newcommand{\dgr}{{$^\circ$}}

\journal{ICARUS}

\begin{document}

\begin{frontmatter}



\title{Characteristics of proton velocity distribution functions in the near-lunar wake from Chandrayaan-1/SWIM observations}




\author[SPL]{M B Dhanya\corref{cor1}} 
\ead{mb\_dhanya\@vssc.gov.in}

\author[SPL]{Anil Bhardwaj} 
\author[IRF]{Yoshifumi Futaana} 
\author[IRF]{Stas Barabash} 
\author[SPL]{Abhinaw Alok} 
\author[IRF]{Martin Wieser} 
\author[IRF]{Mats Holmstr\"{o}m} 
\author[UBe]{Peter Wurz} 

\cortext[cor1]{Corresponding author}
\address[SPL]{Space Physics Laboratory, Vikram Sarabhai Space Center, Trivandrum 695022, India}
\address[IRF]{Swedish Institute of Space Physics, Box 812, Kiruna SE-98128, Sweden}
\address[UBe]{Physikalisches Institut, University of Bern, Sidlerstrasse 5, CH-3012 Bern, Switzerland}


\begin{abstract}
Due to the high absorption of solar wind plasma on the lunar dayside, a large scale wake structure is formed downstream of the Moon. However, recent in-situ observations have revealed the presence of protons in the near-lunar wake  (100 km to 200 km from the surface). The solar wind, either directly or  after interaction with the lunar surface (including magnetic anomalies), is the source of these protons in the near-wake region. Using the entire data from the SWIM sensor of the SARA experiment onboard Chandrayaan-1, we analysed the velocity distribution of the protons observed in the near-lunar wake. The average velocity distribution functions, computed in the solar wind rest frame, were further separated based on the angle between the upstream solar wind velocity and the IMF.   Although the protons enter the wake parallel as well as perpendicular to the IMF, the velocity distribution were not identical for the different IMF orientations, indicating the control of IMF in the proton entry processes. Several proton populations were identified from the velocity distribution and their possible entry mechanism were inferred based on the characteristics of the velocity distribution. These entry mechanisms include (i) diffusion of solar wind protons into the wake along IMF, (ii) the solar wind protons with finite gyro-radii that are aided by the wake boundary electric field, (iii) solar wind protons with gyro-radii larger than lunar radii from the tail of the solar wind velocity distribution, and (iv) scattering of solar wind protons from the dayside lunar surface or from magnetic anomalies.  In order to gain more insight into the entry mechanisms associated with different populations, backtracing is carried out for each of these populations. For most of the populations, the source of the protons obtained from backtracing is found to be in agreement with that inferred from the velocity distribution. There are few populations that could not be explained by the known mechanisms and remain unknown.

\end{abstract}


\end{frontmatter}


\section{Introduction}

When the supersonic solar wind flows past the Moon, it leaves a cavity behind the Moon known as lunar plasma wake. Recent observations in the near-lunar wake region (100 km to 200 km from the surface) have revealed the presence of protons \citep{Nishino09a,Nishino09b,Futaana10b,Wang10,Wiehle11,Dhanya13,Halekas14b}.  These protons, which are basically solar wind protons, reach the near-lunar wake due to various processes \citep{Halekas15a, Bhardwaj15}.

Protons were found to enter the lunar wake parallel to the direction of interplanetary magnetic field (IMF) at 100 km from the lunar surface by SARA/Chandrayaan-1 \citep{Futaana10b}. These observations were at solar zenith angle (SZA) of $\sim$140\dgr\ (50\dgr\ behind the terminator) and the IMF was dominantly perpendicular to solar wind velocity. The energy of the protons was higher compared to that of solar wind and the density was around 0.1\% of the solar wind proton density.  The simple 1-D model for the plasma expansion into the vacuum \citep{Gurevich73,Samir83} did not explain the presence of these protons. In the mid-wake region of $\sim$3.5 R$_L$ downstream (R$_L$ refers to radius of the Moon), ARTEMIS has also observed protons parallel to IMF \citep{Wiehle11, Halekas11b}. These observations were found to be in agreement with the 1-D plasma expansion model at this distance. Recently, \cite{Halekas14b} have investigated the one-dimensional solutions for parallel filling of the wake using hot ion distributions (rather than the conventional cold ion approximation) and using Maxwellian as well as kappa distribution functions for the solar wind electrons. The solutions were compared with ARTEMIS observations and found to be in agreement. 

Several observations have also shown that the  protons enter the near-wake in a direction perpendicular to the IMF. \cite{Nishino09a} have reported observations at 100 km altitude and SZA closer to 150\dgr. These protons from the solar wind enter into the near-lunar wake due to an increase in their gyro-radii under the influence of the electric field at the wake boundary. \cite{Nishino09b} have reported protons at 100 km  altitude closer to SZA of 168\dgr. These protons had a wider energy range: starting from lower than that of solar wind to well above the solar wind energy. The maximum energy observed was around six times that of solar wind. Backtracing of these protons showed that they are solar wind protons scattered at larger angles upon interaction with the dayside lunar surface  \citep{Saito08, Holmstrom10}. Their trajectories under the influence of the IMF and the convective electric field of the solar wind ($E_c$) provide access to the deeper lunar wake. \cite{Wang10} have reported protons in the near-lunar wake closer to the terminator regions of Moon. Closer to the terminator, these protons had energies less than that of solar wind, whereas their energy was found to increase further inside  the lunar wake.  These protons, which travel dominantly perpendicular to IMF, were proposed to be the solar wind protons that are forward scattered from the lunar surface closer to the terminator region. 

In all the above  reported events the background IMF was dominantly perpendicular to the  solar wind velocity (flow).  Later, \cite{Nishino13} showed that the mechanism by which the surface scattered protons reaches the wake \citep{Nishino09b}  operates for other IMF orientations also, except perfectly aligned flow, where the convective electric field, $E_c \sim 0$. \cite{Dhanya13} have reported the proton observation in the low-altitude (100 km) wake  when the IMF was aligned with the solar wind flow, a case where parallel entry along IMF as well as perpendicular entry under the influence of convective electric field cannot play a role.  The energy of the observed protons were higher than that of average solar wind energy by few 100 eV. Those protons are from the higher energy tail of the solar wind velocity distribution. They access the deeper wake due to their larger gyro-radius.

In the context of entry of the protons to the near-lunar wake by various mechanisms, we have investigated the characteristics of velocity distribution of the protons in the low-altitude (100 km to 200 km from the surface) wake by using all the available data from the Solar Wind Monitor (SWIM) sensor of the SARA (Sub-keV Atom Reflecting Analyser) experiment on Chandrayaan-1. This will help us to understand the different proton populations in the lunar wake. 

\section{Instrumentation and Data Sources}

 The solar Wind Monitor (SWIM) was one of the two sensors of the SARA experiment on Chandrayaan-1 mission \citep{Bhardwaj05, Barabash09},  which was a polar orbiting satellite around the Moon. SWIM was an ion-mass analyser with a fan-shaped field of view (FoV). The SWIM FoV was divided into 16 angular pixels of resolution  $\sim 3.5$\dgr\ (elevation)$\times 10$\dgr\ (azimuth), depending on the viewing direction. The energy coverage of SWIM was 100--3000 eV/q with $\triangle$E/E $\sim$7\%. SWIM measured ion flux in logarithmically separated 16 energy steps. The measurement for one energy and direction takes 31.25 ms. The SWIM FoV plane was almost perpendicular to the velocity vector of the spacecraft. One edge of the fan-shaped FoV looks toward the Moon and the other to the zenith. Assisted by the nadir-pointing spacecraft motion, the FoV covers $\sim2\pi$ (half hemisphere) within half of the orbit. However, this means that the SWIM FoV plane changes its orientation according to the latitude. The FoV plane is in the ecliptic plane closer to the equator and it was perpendicular to the ecliptic plane closer to the poles (more details about FoV orientation can be found in Futaana et al., 2010 and Dhanya et al., 2013). 

	Due to the motion of Earth around the Sun, the orbital plane of Chandrayaan~-1 had a drift of
	 $\sim$1\dgr\ per day with respect to the Sun. This helps to have SWIM observations closer to
	  the dawn-dusk terminator (March--April 2009) as well as in the noon-midnight plane (June--July 2009).  Chandrayaan-1 was in a 100 km circular orbit till mid May 2009; subsequently, the
	   orbit was raised to 200 km. The SWIM data obtained from both altitudes (January--July
	    2009 period) are considered for the analysis in this paper.  After excluding the days when
	     the Moon was in the Earth's magnetosheath and magnetotail, about 500 orbits of
	      observation data in total is used in the analysis.

For the upstream solar wind parameters, such as solar wind velocity, density, and IMF orientation, we have used level-2 data from the SWEPAM and MAG instruments on the ACE satellite. Since ACE makes measurements near the L1 point, the data have been time-shifted to the location of Moon by considering the solar wind speed and distance of ACE from the Moon at each instant. 

\section{Co-ordinate systems}

For the analysis, we have used mainly two co-ordinate systems:\\
 (1) aberrated Lunar-centric Solar Ecliptic (aLSE) co-ordinate system, and\\
 (2) Lunar Solar wind Electric field  (LSwE) co-ordinates. \\
  
 In aLSE co-ordinate system, the $x$-axis is towards the anti-solar wind velocity ($-V_{sw}$) which has been corrected for aberration due to the motion of Earth-Moon system, the $z$-axis is towards ecliptic north and $y$-axis completes the right handed co-ordinate system.
 
\begin{figure}
       		   		            \centering
       		   		            \includegraphics[width=30pc]{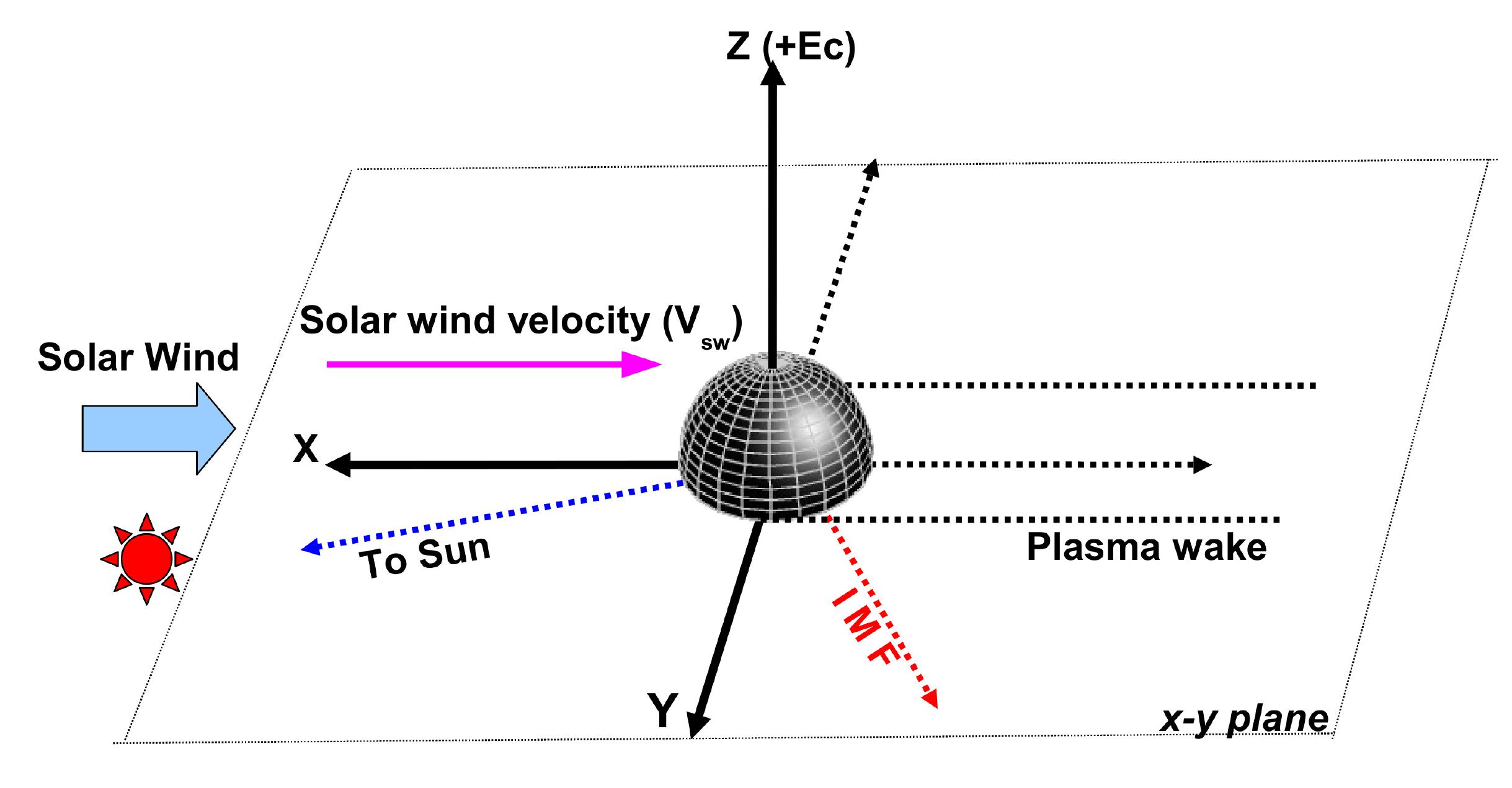}
       		   		          \caption{Illustration of LSwE co-ordinate system. The $x$-axis is along anti-solar wind velocity, $z$-axis is along the convective electric field of the solar wind (${E_c} =-{V_{sw}} \times {B_{IMF}}$) and the $y$-axis completes the right handed co-ordinate system.  In this co-ordinate system  IMF (${B_{IMF}}$) will be confined in the $x$-$y$ plane with the $y$-component always positive (sample direction as red dotted arrow).}
       		   		     	   		\label{fig:lsweframe}
       		   	    \end{figure}

  In LSwE co-ordinates (see Fig.~\ref{fig:lsweframe}), the $x$-axis is towards the anti-solar wind velocity (similar to aLSE), whereas the $z$-axis is along the convective electric field of the solar wind ($E_c =-V_{sw} \times B_{IMF}$), and the $y$-axis completes the right handed co-ordinate system. In this co-ordinates, IMF ($B_{IMF}$) will be confined in the $x$-$y$ plane with the $y$-component always positive. The processes by which the protons enter the near lunar wake may depend on the orientation of IMF and the convective electric field, whose direction varies. The LSwE co-ordinate system helps to confine the direction of IMF in specific plane and that of the convective electric field in particular direction.
  
   In addition, we have used the solar wind rest frame, which is a frame moving with the solar wind and hence the convective electric field is zero. This frame has been used to discuss the velocity distribution function of protons in the wake where proton velocities are resolved into components having direction parallel and perpendicular to the IMF. 

\section{Observation}

\begin{figure}
		        		 		    \includegraphics[width=30pc]{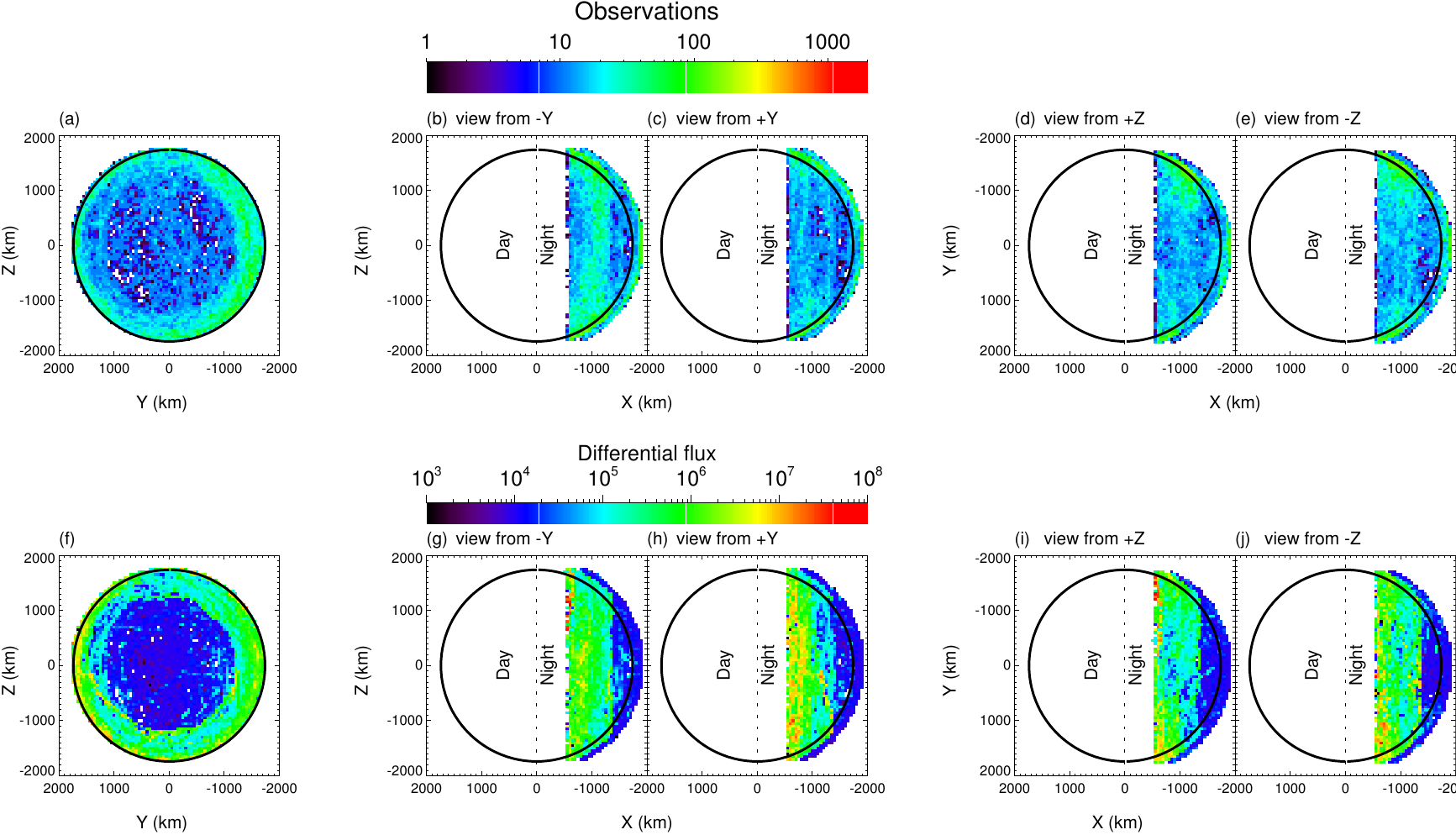}
		        		   		    \caption{Maps of wake protons in LSwE co-ordinate system. The top panel shows the number of observations in different planes whereas the bottom panel shows the observed differential flux. (a) The map of number of observations projected onto the $y-z$ plane, (b) number of observations in the $x-z$ plane ($V_{sw}-E_c$ plane) as viewed from the direction of $-y$, (c) same as (b) but viewed from the direction of $+y$, (d) number of observations in the $x-y$ plane  ($V_{sw}-B_{IMF}$ plane) as viewed from the direction of $+z$ ($+E_c$), (e)  same as (b) but as viewed from the direction of $-z$ ($-E_c$), (f) observed differential flux projected onto the $y-z$ plane, (g) observed differential flux projected onto the $x-z$ plane ($V_{sw}-E_c$ plane) as viewed from the direction of $-y$, (h)same as (g) but as viewed from the direction of $+y$ , (i) observed differential flux projected onto the $x-y$ plane  ($V_{sw}-B_{IMF}$ plane) as viewed from the direction of $+z$ ($+E_c$), (j) same as (i) but as viewed from the direction of $-z$ ($-E_c$).}
		        		   		    
		        		   		   \label{fig:maplsew}
		        		   	\end{figure}

	SWIM made observations in the lunar wake in every orbit, initially  from 100 km altitude, and later from 200 km altitude. The location of the observations in the lunar wake in LSwE (spacecraft location) are binned into grid cells of size 100 km $\times$ 100 km.  The differential flux observed by SWIM when the spacecraft was located over any grid cell is accumulated in that grid cell.  Since there will be multiple observations over any grid cell, due to repeated orbits, the accumulated counts on a given grid cell is divided by the number of observations in that grid cell to get the average differential flux. The 2D maps of the number of observations in the lunar wake and the observed differential flux, projected in   different planes of the LSwE co-ordinates are shown in Fig.~\ref{fig:maplsew}.
	
\begin{figure}
           		   \centering
           		 		     \includegraphics[width=30pc]{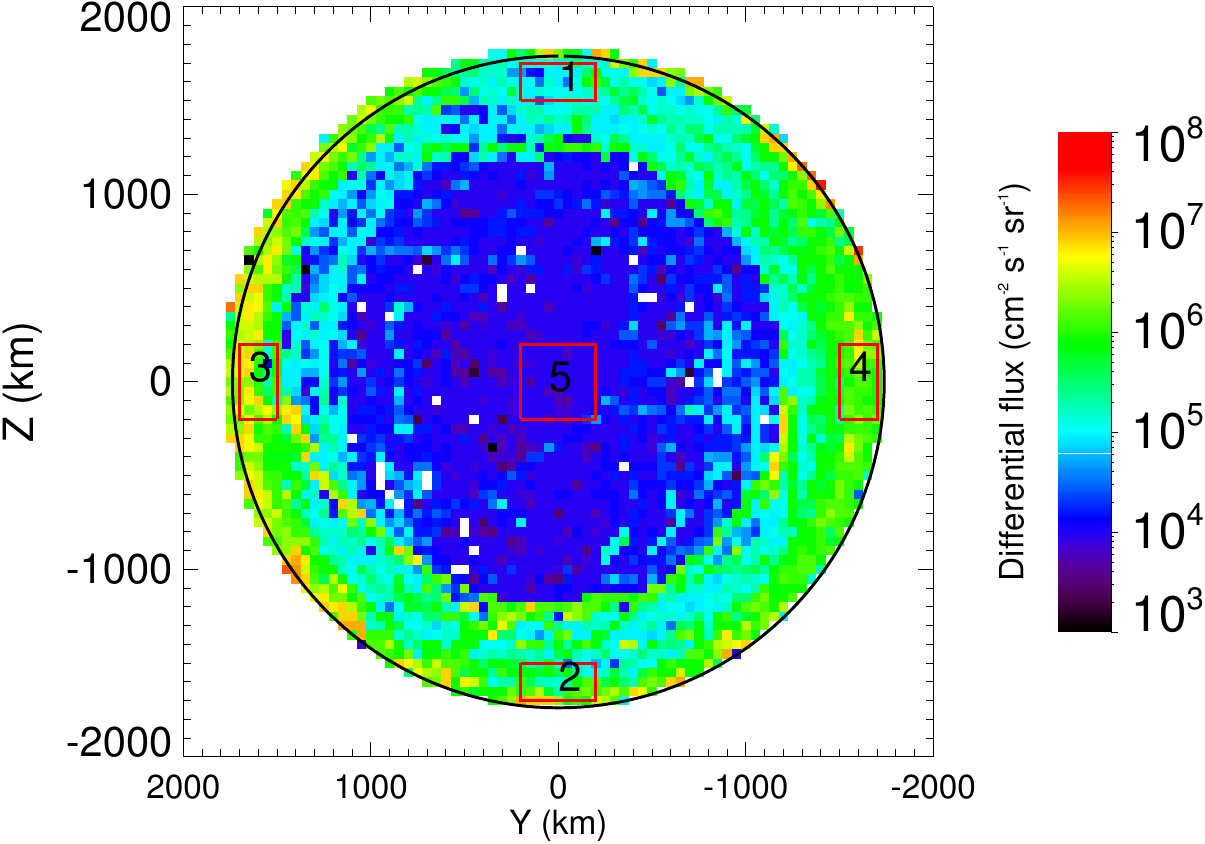}
           		   		    \caption{Map of differential flux of wake ions in the $y-z$ plane of the LSwE co-ordinate system, showing the 5 locations of interest (enclosed by red rectangular boxes). Location 1 at $+E_c$ limb, location 2 at $-E_c$ limb, location 3 at $+y$ limb, and location 4 at $-y$ limb, and location 5 at the central wake. The locations 3 and 4 are in the plane containing solar wind velocity and IMF.}
           		   		   \label{fig:regions}
           \end{figure}

	The top panels (Fig.~\ref{fig:maplsew}a to Fig.~\ref{fig:maplsew}e) show the 2-D map of number of observations projected onto  different planes. In most cases, there are several observations over a given location. The corresponding differential flux  has an asymmetry  along the $-z$-axis ($E_c$), with higher differential flux at the $+E_c$ ($+z$) hemisphere compared to the $-E_c$ ($-z$) hemisphere within few 100 kms from the terminator (strip in red color seen Fig.~\ref{fig:maplsew}i compared to Fig.~\ref{fig:maplsew}j). This asymmetry may be due to the role of the electric field in aiding the entry of protons to near-wake across the $+E_c$ pole. Similarly, the differential flux is asymmetric along $y$, with the flux in $-y$ (Fig.~\ref{fig:maplsew}g) being higher than in $+y$  (Fig.~\ref{fig:maplsew}h), closer to limb.  Since IMF lies in the $x$-$y$ plane with $y$-component always positive,  the asymmetry between $-y$ and $+y$ may be associated with the diffusion of solar wind parallel ($-y$) and anti-parallel to IMF ($+y$). Hence, for further analysis we have chosen five locations as shown in Fig.~\ref{fig:regions}. The location 1 is closer to $+E_c$ pole, location 2 is closer to $-E_c$ pole, location 3 is closer to $+y$ limb, location 4 is closer to $-y$ limb, and location 5 is in the central wake. While the location 5 consists of observations from spatial grid extend of 400 km $\times$ 400 km (in $y$-$z$ plane in LSwE co-ordinates), other locations are rectangular having spatial extents of 200 km  $\times$ 400 km. The locations 1 to 4 are at a distance of 1500 km to 1700 km from the centre of the wake. Due to the finite gyro-radius of protons in the solar wind, there is always a possibility for protons to be found in lunar wake close to the terminator.  Hence the observations very close to the terminator (limb observations) are excluded.
	
     Using the information on the energy and the direction of the particles observed by SWIM in the above mentioned five locations, the velocity distribution functions are computed in the aLSE co-ordinates and transformed to the solar wind rest frame. The velocity distribution functions are indeed functions of three dimensional velocity space, but to understand the population in a simple manner, we used the two dimensional velocity space, namely in  the directions parallel and perpendicular to the IMF ($v_\parallel$  and  $v_\perp$). Note that we employ the solar wind rest frame for this analysis. Indeed this frame is frequently used for the pitch angle (=arctan($v_\perp$/$v_\parallel$)) analysis. This frame is helpful for understanding the physical mechanisms of acceleration and the initial velocity vector of the protons, if they are produced near the Moon (such as exospheric ions).  
    
\begin{figure}
                        		   \centering
                        		 		     \includegraphics[width=30pc]{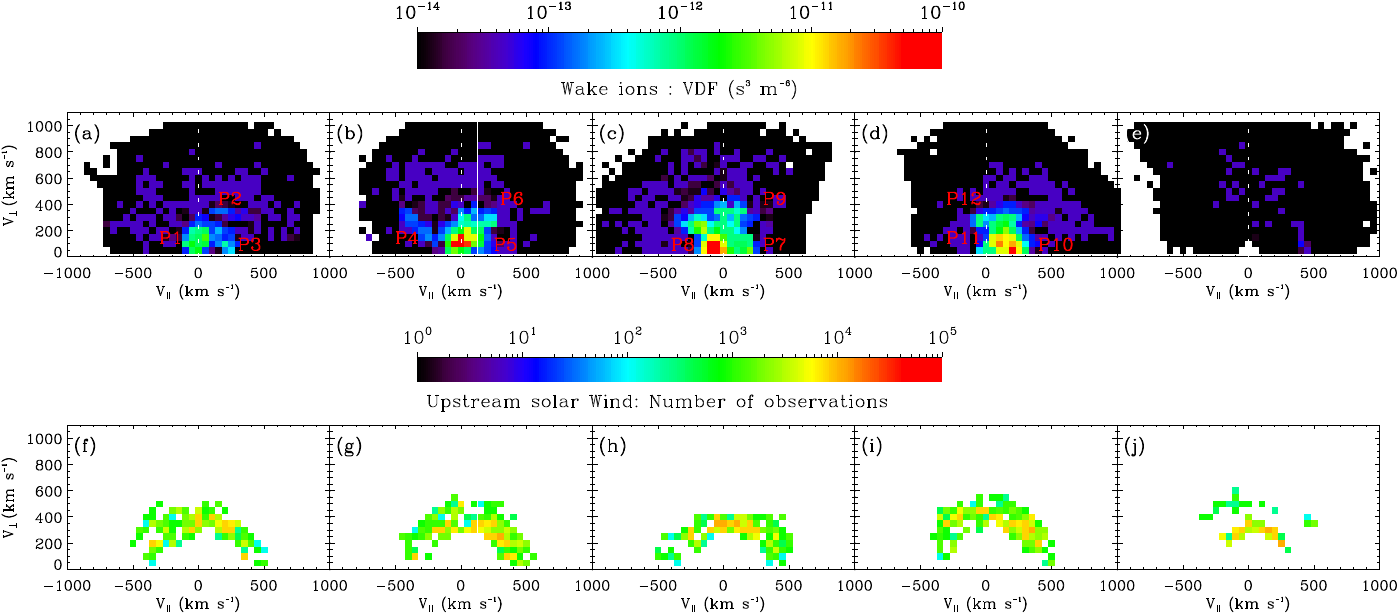}
                        		   		    \caption{ Top panel shows the velocity distribution of protons in the near lunar wake in the solar wind rest frame, and bottom panel shows the upstream solar wind distribution (from ACE) corresponding to the SWIM observations at each of the locations 1 to 5 . (a) velocity distribution of protons observed in location 1, (b) velocity distribution of protons observed in location 2, (c) velocity distribution of protons observed in location 3, (d) velocity distribution of protons observed in location 4, (e) velocity distribution of protons observed in location 5, (f) velocity distribution of upstream solar wind during the SWIM observations in location 1, (g) velocity distribution of upstream solar wind during the SWIM observations in lunar wake in location 2, (h)  velocity distribution of upstream solar wind during the SWIM observations  in lunar wake in location 3, (i) velocity distribution of upstream solar wind during the SWIM observations  in lunar wake in location 4, (j) velocity distribution of upstream solar wind during the SWIM observations  in lunar wake in location 5. Refer to Fig.~\ref{fig:regions} for the definition of the different locations.  `P1' to `P12' indicate the prominent proton populations at the different locations.
                        		   		    }
                        		   		    
                        		   		   \label{fig:veldist1}
            \end{figure}

    To understand the prevailing upstream solar wind conditions during the SWIM observations at locations 1 to 5 in lunar wake, the  velocity distribution of the upstream solar wind was also computed. The solar wind velocity data from the ACE spacecraft that has been time shifted to the location of Moon was  resolved into components parallel ($v_\parallel$) and perpendicular ($v_\perp$) to the IMF in aLSE co-ordinates.  The velocity components ($v_\parallel$  and  $v_\perp$) are binned into 50 km s$^{-1}$ $\times$ 50 km s$^{-1}$ grids. The frequency distribution, i.e., how many times the solar wind velocity was observed to be in a specific bin, for those days and timings where SWIM had observations over location-1 to location-5 is considered as the representative of the upstream solar wind condition. Fig.~\ref{fig:veldist1} shows the velocity distribution of the protons in the solar wind rest frame that are observed in locations 1 to 5 together with the upstream solar wind distribution in aLSE co-ordinate system. 
    
     From Fig.~\ref{fig:veldist1}a--\ref{fig:veldist1}e, it can be seen that all the locations except location-5, have several velocity bins with significant values of  the distribution function  above the prevailing background.  The velocity bins which have value of the distribution function above 10$^{-13}$ s$^3$ m$^{-6}$ are considered to be significant above the background (1 count level). To identify the different populations of protons from the velocity distribution map, a selection criterium was employed in which any velocity bin is analysed in the following manner.  In the analysis, all the eight velocity bins surrounding a given bin (say center bin) are considered. Please note that all the bins referred here are in the velocity space ($v_\parallel$, $v_\perp$). The value of the distribution function in the center bin is compared with that of the surrounding eight bins and if the value in the surrounding bins are lower than that of the center bin by 50\%, the proton velocity distribution represented in the center bin together with the surrounding bins are considered as belonging to the same population.
     
    If the value of the distribution function in any of the eight bins (say $vbin1$) is higher than that of the center bin by more than 50\%, then $vbin1$ is considered as the new center bin. The value of the distribution function in the eight bins surrounding $vbin1$ are compared with that of $vbin1$. This continues till all the eight velocity bins surrounding a center bin are found to have value less than that of the center bin (by 50\%). Once this is met, the velocity distribution represented from the first center bin till the final center bin are considered as belonging to a given population. According to this selection criterium, the prominent populations are marked in Fig.~\ref{fig:veldist1} (P1 to P12). For location 5, which is close to the central wake region, the signal strength is too low for any distinct proton population to be seen. 
    
    In location 1 (Fig.~\ref{fig:veldist1}a), protons of population P1 have the values of ($v_\parallel\ ,v_\perp$) around (0, 150) km s$^{-1}$, those of P2 around (150, 350) km s$^{-1}$, and of P3  around (250, 50) km s$^{-1}$.   Comparison with the corresponding average upstream solar wind condition (Fig.~\ref{fig:veldist1}f) shows that protons belonging to P1 and P3 have velocities lower than that of solar wind ($\sim$400 km s$^{-1}$), whereas P2 have velocity comparable to that of solar wind. P1 comprises of protons of dominant  $v_\perp$ with negligible $v_\parallel$, P2 have $v_\perp$ comparable to that of solar wind with non-zero $v_\parallel$, and P3  have dominant $v_\parallel$ with negligible $v_\perp$. 
    
    In location 2 (Fig.~\ref{fig:veldist1}b), three prominent proton populations are observed (P4 to P6). Protons belonging to P4 that have comparable $v_\parallel$ and $v_\perp$ appear as a  strip in the velocity space with $v_\parallel$ in the range $-200$ to $-450$  km s$^{-1}$ and $v_\perp$ in the range 150 to 350  km s$^{-1}$. P5 is centered around velocity bins of (0, 150) km s$^{-1}$, and P6 have $v_\parallel$ in the range 150 to 300  km s$^{-1}$ and $v_\perp$ in the range 200 to 400  km s$^{-1}$. Comparison with the corresponding upstream solar wind condition (Fig.~\ref{fig:veldist1}g) shows that while P4, and P6 may have velocities comparable to that of solar wind, P5 have velocities lower than that of solar wind.
    
     Three populations (P7 to P9) are observed in location 3 (Fig.~\ref{fig:veldist1}c) of which P7 have velocities centered around (200, 50) km s$^{-1}$,  P8 around ($-$50, 100) km s$^{-1}$, and P9  around (100, 350) km s$^{-1}$. Populations P7 and P8 have  velocities less than that of the corresponding upstream solar wind (Fig.~\ref{fig:veldist1}h), whereas P9 have velocities comparable to that of solar wind. In location 4 (Fig.~\ref{fig:veldist1}d), three populations (P10 to P12) are identified. Population P10 is centered around (200, 50) km s$^{-1}$, P11 around (100, 100) km s$^{-1}$, and P12 around ($-$50, 300) km s$^{-1}$. All of them have velocities lower than that of the upstream solar wind (Fig.~\ref{fig:veldist1}i).

\begin{figure}
       	               		   \centering
       	               		 		     \includegraphics[width=30pc]{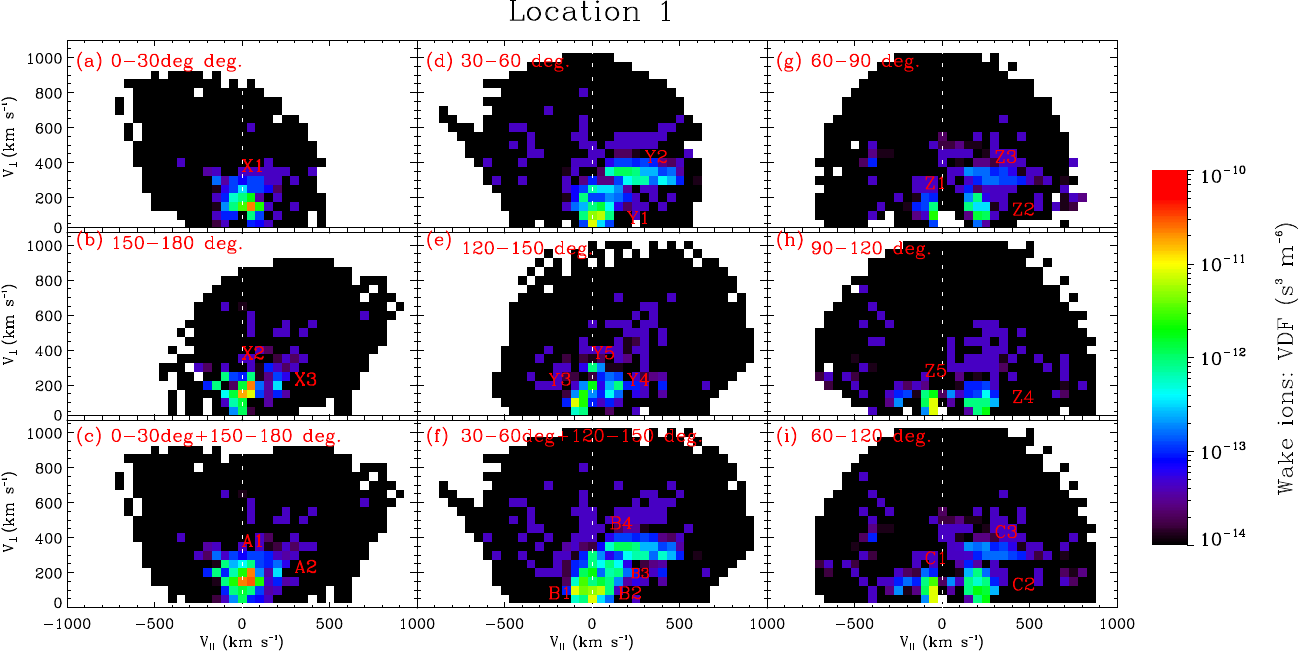}
       	               		   		    \caption{ The velocity distribution of wake protons at location 1 in the solar wind rest frame under different IMF angles of (a) 0\dgr--30\dgr, (b) 150\dgr--180\dgr, (c) 0\dgr--30\dgr\ and 150\dgr--180\dgr\ combined, (d) 30\dgr--60\dgr, (e) 120\dgr--150\dgr, (f) 30\dgr--60\dgr\ and 120\dgr--150\dgr\ combined, (g) 60\dgr--90\dgr, (h) 90\dgr--120\dgr, (i) 60\dgr--120\dgr\ combined bin. The white colored dotted vertical line in each panel is drawn at $v_\parallel= 0$. The prominent proton populations are marked in  the figure (X1, X2, X3, A1, A2, Y1, Y2, Y3, Y4, Y5, B1, B2, B3, B4, Z1, Z2, Z3, Z4, Z5, C1, C2, C3). Summary of the velocity components of the populations along with that of the  upstream solar wind  can be found in Table~\ref{table:VDF}.}
       	               		   		    
       	               		   		   \label{fig:veldist2}
       	               \end{figure}

    Further, the velocity distribution were separated based on the angle between the solar wind velocity and the interplanetary magnetic field in aLSE ($\theta_{IMF}$). Bins of 30\dgr\ with $\theta_{IMF}$ in the range 0\dgr\ to 30\dgr, 150\dgr\ to 180\dgr, 30\dgr\ to 60\dgr, 120\dgr\ to 150\dgr, 60\dgr\ to 90\dgr\ and 90\dgr\ to 120\dgr, are considered. Since there was no significant proton population in location 5 (Fig.~\ref{fig:veldist1}e), location 5 was not included in the further analysis. The IMF sorted velocity distribution for location 1 is shown in Fig.~\ref{fig:veldist2}. For  comparison, the distribution for the combined $\theta_{IMF}$ bins where  0\dgr\ to 30\dgr\ and 150\dgr\ to 180\dgr\ are combined, 30\dgr\ to 60\dgr\ and 120\dgr\ to 150\dgr\ combined, 60\dgr\ to 90\dgr\ and 90\dgr\ to 120\dgr\ combined are also shown in the bottom panel of Fig.~\ref{fig:veldist2}. The prominent proton populations are marked in the figure.
    
   	The proton populations P1 to P3, seen in Fig.~\ref{fig:veldist1}a, show up for different bins of $\theta_{IMF}$ as seen from the Fig.~\ref{fig:veldist2}. X1 (Fig.~\ref{fig:veldist2}a) and X2  (Fig.~\ref{fig:veldist2}b) have similar velocity components centred around (50, 150) km s$^{-1}$ and clearly represent the distribution A1 (Fig.~\ref{fig:veldist2}c) for the combined $\theta_{IMF}$ of 0\dgr\ to 30\dgr\ and 150\dgr\ to 180\dgr. The population X3 is similar to A2. Similarly, for the $\theta_{IMF}$ of 30\dgr\ to 60\dgr\ and 120\dgr\ to 150\dgr, the distributions Y1 (Fig.~\ref{fig:veldist2}d) contributes to B2 (Fig.~\ref{fig:veldist2}f), Y2 (Fig.~\ref{fig:veldist2}d) contributes to B4 (Fig.~\ref{fig:veldist2}f), Y3 (Fig.~\ref{fig:veldist2}e) contributes to B1 (Fig.~\ref{fig:veldist2}f), Y4 (Fig.~\ref{fig:veldist2}e) contributes to B3 (Fig.~\ref{fig:veldist2}f),  and Y5 (Fig.~\ref{fig:veldist2}e) is also part of B3 (Fig.~\ref{fig:veldist2}f). Similar is the situation for the $\theta_{IMF}$ of 60\dgr\ to 90\dgr\ and 90\dgr\ to 120\dgr\ where C1  (Fig.~\ref{fig:veldist2}i) is contributed from Z1  (Fig.~\ref{fig:veldist2}g) and Z5  (Fig.~\ref{fig:veldist2}h); C3 (Fig.~\ref{fig:veldist2}i) from Z3 (Fig.~\ref{fig:veldist2}g); and C2 (Fig.~\ref{fig:veldist2}i) from Z2 (Fig.~\ref{fig:veldist2}g) and Z4  (Fig.~\ref{fig:veldist2}h).
   	
   	 \begin{figure}
   	   	            		   \centering
   	   	            		 		     \includegraphics[width=30pc]{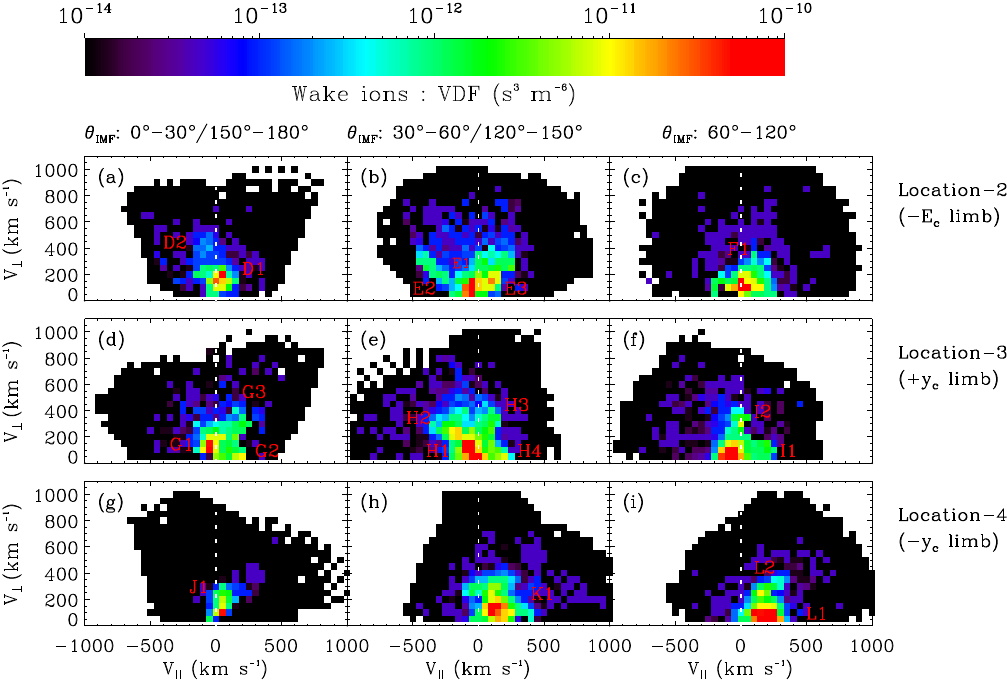}
   	   	            		   		    \caption{ The velocity distribution of protons in lunar wake in the solar wind rest frame for location 2 to location 4 under different values of $\theta_{IMF}$ (angle between solar wind velocity and IMF in aLSE co-ordinates). Each horizontal panel represents the distribution for a given location as indicated on the extreme right of the panels. Within a panel, each plot from left to right represents the velocity distribution for different bins of $\theta_{IMF}$, which is indicated on the top. (a) distribution for $\theta_{IMF}$ bin 1 at  location 2, (b) for $\theta_{IMF}$ bin 2 at location 2, (c) for $\theta_{IMF}$ bin 3 at location 2, (d) for $\theta_{IMF}$ bin 1 at  location 3,  (e) for $\theta_{IMF}$ bin 2 at location 3, (f) for $\theta_{IMF}$ bin 3 at  location 3, (f) for $\theta_{IMF}$ bin 1 at  location 4, (g) for $\theta_{IMF}$ bin 2 at  location 4, (h) for $\theta_{IMF}$ bin 3 at  location 4 (see Fig.~\ref{fig:regions} for the definition of the different locations). The $\theta_{IMF}$ bin 1 refers to 0\dgr--30\dgr\ and 150\dgr--180\dgr\ combined, $\theta_{IMF}$ bin 2 refers to 30\dgr--60\dgr\ and 120\dgr--150\dgr\  combined,  $\theta_{IMF}$ bin 3 refers to 90\dgr--120\dgr. The white colored dotted vertical line in each panel is drawn at v$_\parallel= 0$. The prominent proton populations are marked in red color from `D2' to `L2' and their velocity components  along with that of upstream solar wind are given in  Table~\ref{table:VDF}.   }
   	   	            		   		    
   	   	            		   		   \label{fig:veldist3}
   	   	            \end{figure}
   	   	            
   	Similar scenario was found to hold for location 2 to location 4 (not shown). Hence, the proton distribution for only the combined bins of $\theta_{IMF}$ are presented for locations 2 to 4 (Fig.~\ref{fig:veldist3}). The prominent proton populations identified  in each of the locations based on the selection criteria are marked in the Fig.~\ref{fig:veldist3} from D2 to L2.  A summary of the prominent proton populations in each of the locations 1 to 4 and their velocity components are given in Table~\ref{table:VDF}.


   	            \begin{table}
   	            \caption{Proton velocity distribution in location 1 to location 4 } 
   	            \label{table:VDF}
   	            \centering
   	            \begin{tabular}{cccccc}
   	            \hline
   	            Location & Population & $v_\parallel^a$ \  &  $v_\perp^b$ \  & $|v_p|^c$\   & $|V_{sw}|^d$\\
   	            \hline
   	             1 & A1  & 0 & 150  & 150.0  & 400 \\
   	             1 & A2  & 200 & 200  & 282.8 & 400 \\
   	             1 & B1  & -100 & 100   & 141.4 & 400\\
   	             1 & B2  & 0 & 200   & 200.0 & 400\\
   	             1 & B3  & 150 & 200   & 250.0 & 400\\
   	             1 & B4  & 200 & 350   & 403.1 & 400\\
   	             1 & C1  & -50 & 100  & 111.8 & 400\\
   	             1 & C2  & 200 & 100  & 223.6 & 400\\
   	             1 & C3  & 200 & 350  & 403.1 & 400\\
   	             2 &  D1  & 50 & 200  & 206.1 & 350 \\
   	             2 &  D2  & -50 & 450  & 452.8 & 350 \\ 
   	             2 &   E1  & -50 & 100 & 111.8  & 350\\
   	             2 &   E2  & -350 & 250 & 430.11  & 350\\
   	             2 &   E3  & 100 & 150 & 180.3  & 350\\  
   	             2 & F1  & 0 & 100  & 100.0 & 350 \\
   	             3 & G1  & -50 & 100    & 111.8 & 350 \\
   	             3 & G2  & 200 & 50    & 206.1 & 350 \\
   	             3 & G3  & 200 & 350    & 403.1 & 350 \\  
   	             3 &  H1  & 0 & 200  & 200.0 & 350 \\
   	             3 &  H2  & -150 & 250  & 291.5 & 350 \\
   	             3 &  H3  & 100 & 350  & 364.0 & 350 \\
   	             3 &  H4  & 200 & 50  & 206.1 & 350 \\ 
   	             3 &  I1  & -100 & 50  & 141.4 & 350 \\
   	             3 &  I2  & 0 & 300  & 300.0 & 350 \\
   	             4 &  J1  & 50 & 100  & 111.8 & 450 \\
   	             4 &  K1  & 150 & 150    & 212.1 & 450 \\
   	             4 &  L1   & 200 & 50    & 250  & 450 \\
   	             4 &  L2  & 150 & 300  & 335.4 & 450 \\
   	             
   	             \hline
   	            \end{tabular} \\
   	            \centering {\small All velocity values in  km s$^{-1}$} \\
   	            \flushleft {$^a$\small velocity of wake protons parallel to IMF in the solar wind rest frame }
   	            \flushleft {$^b$\small velocity of wake protons perpendicular to IMF in the solar wind rest frame }\\
   	            \flushleft {$^c$\small velocity magnitude of wake protons in solar wind rest frame }\\
   	            \flushleft {$^d$\small solar wind velocity in aLSE co-ordinates }
   	            
   	            \end{table}               
    
\section{Discussion}

		 The source of protons in the near-lunar wake can be categorized into two. The first is the direct solar wind entry \citep{Nishino09a, Futaana10b, Dhanya13}, and the second is the solar wind after modification due to interaction with lunar surface and magnetic anomalies \citep{Nishino09b, Wang10}. The term `direct' is used to indicate the absence of interaction with lunar surface or magnetic anomalies. The processes which are known for the direct entry are: (1) protons diffusing along the IMF (parallel or anti-parallel to IMF) which can be termed as parallel entry \citep{Futaana10b},  (2) increase in Larmor radii due to the wake boundary electric field \citep{Nishino09a}, and (3) protons from the tail of the solar wind velocity distribution having large gyro-radii \citep{Dhanya13}. The scattering of the solar wind upon interaction with the lunar surface or magnetic anomalies and their subsequent trajectory under IMF and convective electric field enable the protons of the second category to enter the near lunar wake \citep{Nishino09b, Wang10}.

		In the solar wind rest frame, the parallel entry protons will have dominant $v_\parallel$ with almost zero or much smaller $v_\perp$.   The magnitude of the velocity ($v_p$) may be less than that of solar wind (solar wind speed in aLSE). It is to be noted that although \cite{Futaana10b} discussed the observation when IMF was perpendicular to the solar wind velocity vector ($\theta_{IMF} \sim 90$\dgr), this mechanism can operate for other IMF orientations as well, except for magnetic aligned flow ($\theta_{IMF} \sim 0$\dgr).  For convenience, we use the term `Entry-1' for this process, in further discussion. Let's recall that in the LSwE co-ordinates, the IMF is confined in the $x$-$y$ plane with $y$-component always positive and V$_{sw}$ is along $-x$. Hence, for IMF angles of 0\dgr\ to 30\dgr, and 150\dgr\ to 180\dgr, these protons may not be able to reach location 1, which is closer to $+E_c$ ($+z$) pole. This applies for location 2 also, which is closer to  $-E_c$ ($-z$) pole. The parallel entry protons will be able to reach location 3 (along $+y$) and location 4 (along $-y$) for all IMF angles except  $\theta_{IMF} \sim$ 0\dgr. The populations `G2' and `H4' in location 3, with dominant $v_\parallel$ and negligible $v_\perp$ are most likely due to Entry-1. The population `I1' in location 3 and `L1' in location 4 may also be associated with  Entry-1.
		
		
		When the solar wind protons enter the lunar wake due to their finite gyro-radius aided by the wake boundary electric field, they travel in a direction perpendicular to IMF \citep{Nishino09a}. In the solar wind rest frame, these protons are expected to have a dominant $v_\perp$ component and may have a small  component of thermal velocity parallel to the magnetic field. Due to the wake boundary electric field, $v_\perp$ will be enhanced above the thermal velocity. These protons may be characterized with $v_\perp > v_\parallel$, and since $v_\parallel$ is expected to be negligible, it can be considered that $v_\parallel$ $< 0.25 \times v_\perp$. The velocity of such protons is expected to be lower than the solar wind bulk velocity. Although this mechanism has been reported for $\theta_{IMF} \sim 90$\dgr, this mechanism is likely to be active for other IMF orientations. For convenience, we use the term `Entry-2' for this process in further discussions. `Entry-2' is same as `Type-I', defined in \cite{Nishino09b} for the mechanism described in \cite{Nishino09a}. The protons of Entry-2 can reach the $-E_c$ and $+E_c$ poles (locations 1 and 2) for 0\dgr\ $\le \theta_{IMF} \le$ 90\dgr. These protons may access locations 3 and 4 for smaller values of $\theta_{IMF}$ ($\sim$ 0\dgr), whereas for larger values of $\theta_{IMF}$, these protons will start to diffuse along the IMF into the wake. Also, Entry-2 protons may not access the central wake, i.e., location 5. The populations A1, B2, C1 in location 1; D1, E1, F1 in location 2; and G1, H1 in location 3 have the  characteristics of velocity distribution to be associated with Entry-2. 
		
	    
	     The protons which enter the lunar wake by virtue of their large  gyro-radius  \citep{Dhanya13} from the high-energy tail of solar wind velocity distribution (having large thermal velocities) are expected to have dominant $v_\perp$. Such protons have gyro-radii larger than the lunar radius in the solar wind. Since the gyro-radius depends on the IMF, which is a variable, the perpendicular velocity ($v_\perp$) of these protons can be either higher or lower than that of solar wind bulk speed depending on the strength of IMF and the magnitude of V$_{sw}$. For a typical solar wind condition of 400 km s$^{-1}$ bulk speed and 5 nT IMF, the value of $v_\perp$ could be be higher than V$_{sw}$ for the protons to have gyro-radii larger than lunar radius. If IMF is weak, say 1 nT, then for a smaller $v_\perp$ itself  ($\sim$200 km s$^{-1}$), the gyro-radius of protons can be  larger than lunar radius. Since these protons have larger thermal velocity, they may have significant $v_\parallel$ component also. Although this mechanism have been observed during magnetic field aligned flow ($\theta_{IMF} \sim 0$\dgr), this can be active for other IMF orientations as well, although the flux may be lower. We use the term `Entry-3' for this process. Since protons of Entry-3 depend solely on their gyro-radii to enter into the wake, they may access all the five locations. The populations A1, A2, B1, B2, B3, C2 in location 1; D1, D2 , E1, E3, F1 in location 2; G1, G3, H1, H2, H3, I1, I2 in location 4; and J1, K1, L2 in location 4 have the characteristics of Entry-3 protons.  To assess the contribution of Entry-3 in the above populations, we considered the magnitude of IMF that was prevailing during the observation. From the IMF magnitude, the minimum  $v_\perp$ that is required for the gyro-radius of protons to be equal to that of lunar radius, is estimated. This estimated $v_\perp$ is compared against the observed $v_\perp$ of different populations. This analysis showed that Entry-3 can play a role for the populations B2, B3 in location 1; D2, E1, E3, F1 in location 2; G1, G3, H1, H2, I2 in location 3; and J1, K1 in  location 4. Entry-3 may contribute partially for C2, and L2 because of the low IMF condition during some of the observation of these populations.
	     
		  
		It has been observed that  0.1--1\% of the solar wind protons scatter back to space upon interaction with the lunar surface \citep{Saito08}. Also, $<$50\% of solar wind protons scatter upon interaction with the lunar magnetic anomalies \citep{Saito10, Lue11}. The protons scattered from anomalies are found to have velocities almost similar to that of solar wind but heated up \citep{Saito10, Lue11}. In solar wind rest frame, such protons will have similar velocity distribution as that of surface scattered protons. The trajectories of these protons would be controlled by IMF and the convective electric field \citep{Holmstrom10} and some of which can eventually access the near-lunar wake \citep{Nishino09b, Wang10}. We refer to such an entry mechanism as `Entry-4'. Although Entry-4 is similar to the Type-II mechanism reported by \cite{Nishino09b} with reference to the surface scattered protons, Entry-4 includes protons scattered by magnetic anomalies as well. In the solar wind frame, the backscattered protons will have velocities in the range $V_{sw}$ to $2\times V_{sw}$ considering the extreme scattering angles of 90\dgr\ and 180\dgr, and assuming negligible energy loss upon scattering.  In the Moon reference frame, since the backscattered protons access the wake by travelling along $+E_c$ and crossing the $+E_c$ terminator \citep{Nishino09b}, they are not likely to be seen in location 2, whereas they can access location 3 and location 4 for oblique IMF orientations.  The populations B4, and C3 in location 1 with dominant $v_\perp$ and velocities comparable to that of the background solar wind are most likely due to Entry-4. The population G3, H3, I2 in location 3 also have the characteristics associated with  Entry-4. The population D2 in location 2 also have the characteristics of Entry-4 protons, but the location is close to $-E_c$ pole and hence Entry-4 is not expected.

	    The protons which are forward scattered at the terminator  would also be able to reach the lunar wake region \citep{Wang10}, where forward scattered protons closer to the $-E_c$ pole enter the nightside.  We use the term `Entry-5' for such a process. Entry-5 is used here for the forward scattered protons close to the terminator in any region (not only $-E_c$ pole). In the solar wind rest frame, the forward scattered protons will have velocities in the ranging from zero to $V_{sw}$ for the two extreme scattering angles 0\dgr\ and 90\dgr, assuming negligible energy loss upon scattering. Also, such protons are expected to have dominant $v_\perp$ as seen from the solar wind frame. Since the forward scattered protons can access the wake across $-E_c$ terminator \citep{Wang10}, they can be seen in location 2. The populations A1, A2, B1, B2, B3, C1 in location 1; D1, E1, E3, F1 in location 2; G1, H1, H2, I1 in location 3; and J1, K1, L2 in location 4 have the characteristics to be associated with Entry -5.

	  	The summary of the various entry mechanisms discussed above, the expected velocity distribution of protons in solar wind rest frame for each mechanism and the observed populations which are most likely associated with each of the entry mechanisms are provided in Table~\ref{table:summary}. It is seen that for several populations, more than one entry mechanism may play a role.
		
		\begin{table}
		\caption{Mechanisms of entry of solar wind protons in the near-lunar wake,  expected velocity distribution in solar wind rest frame, and the observed populations which are most likely associated with each of the entry mechanisms}
		\label{table:summary}
		\centering
		\begin{tabular}{llll}
		\hline
		 Process  & References & Expected velocity distribution & observed population \\
		 \hline
		 Entry-1 & \cite{Futaana10b}  & Dominant $v_\parallel$ ($v_\perp \sim$ 0); $v_{p} << V_{sw}$ & G2, H4, I1, L1  \\[5pt]
		 
		 Entry-2 & \cite{Nishino09a}  & Dominant $v_\perp$ ($v_\parallel \sim$ 0); $v_{p} < V_{sw}$  & A1, B2, C1, D1, E1, F1, \\
		 & & & G1, H1 \\[5pt] 												
		 Entry-3 & \cite{Dhanya13}  & Dominant $v_\perp$; $v_{p}$ may be greater  & B2, B3, D2, E1, E3, F1, \\
		 
		 &  & or smaller than $V_{sw}$ & G1, G3,
		 H1, H2, I2, J1, \\
		& & & K1, C2*, L2*  \\[5pt]
		 Entry-4 & \cite{Nishino09b},  & Dominant $v_\perp$ ; $v_{p}: \sim V_{sw} \ to\  2* V_{sw}$ & B4, C3, G3, H3, I2 \\
		 Entry-5 & \cite{Wang10}     & Dominant $v_\perp$; $v_{p} << V_{sw}$ &  A1, A2, B1, B2,
		 B3, C1, \\
		 & & & D1, E1, E3, F1, G1, H1, \\
		 & & & H2, J1, K1, L2 \\[5pt]
		 \hline
		\end{tabular}
		\flushleft {* Partially due to the low IMF condition during some of the observation of these populations \\
		$v_{p}$ =  velocity of the wake protons \\
		V$_{sw}$ = solar wind bulk velocity}
		\end{table}
		
\begin{figure}
		                         		   \centering
		                         		 		     \includegraphics[width=25pc]{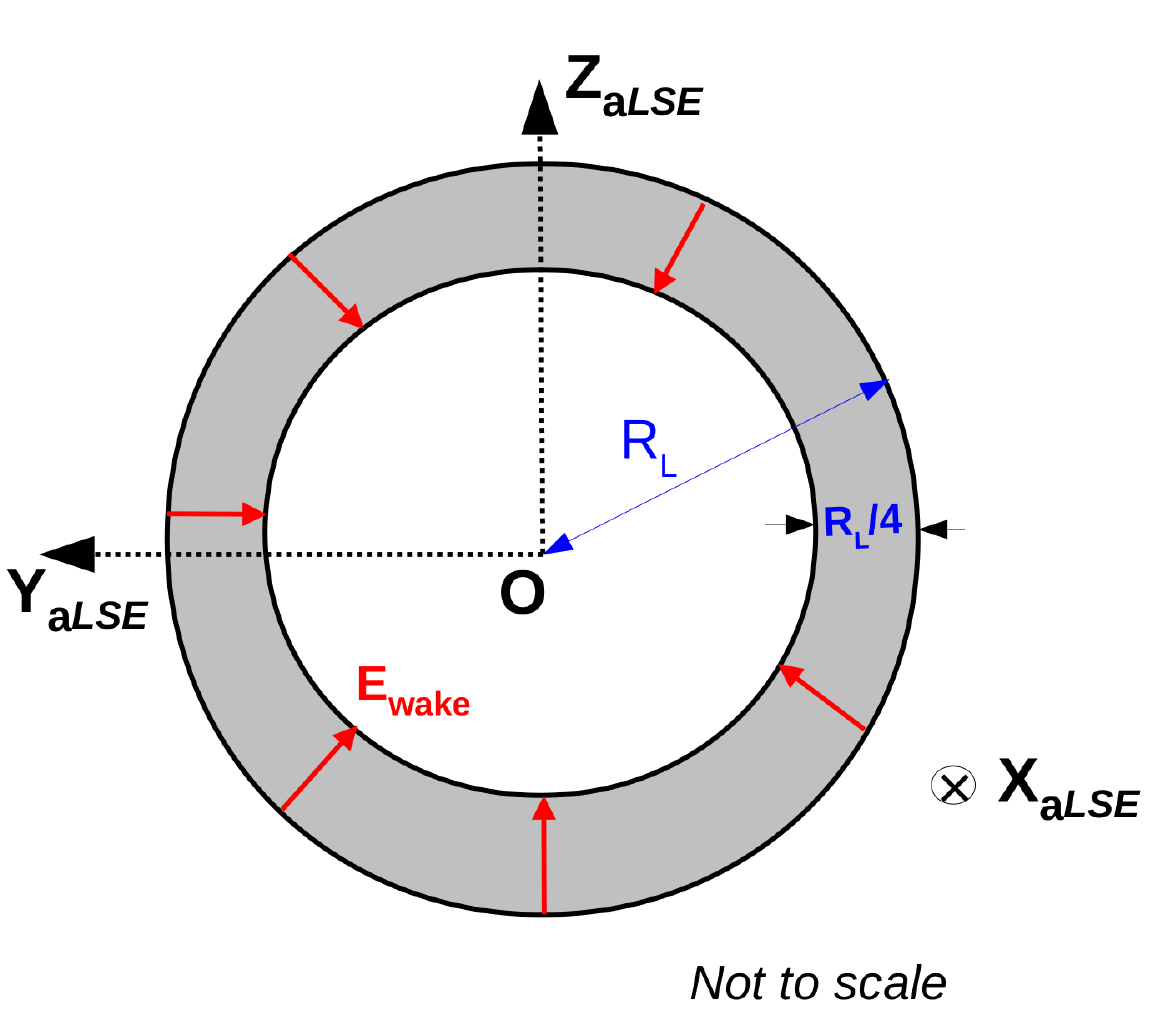}
		                         		   		    \caption{ Illustration of the lunar wake boundary electric field (E$_{wake}$) taken in the proton backtracing model. The outer circle represents the lunar terminator (wake boundary). E$_{wake}$ is shown by red arrows from the wake boundary pointing inwards towards the Sun-Moon line. E$_{wake}$  acts in the shaded region of width  $0.25 \times R_L$ from the wake boundary, where R$_L$ refers to the lunar radius. The inner circle has a radius of 0.75$\times$ R$_L$.   The magnitude of E$_{wake}$ is 0.7 mV m$^{-1}$. This scheme is similar to that discussed in \cite{Nishino09a} and agrees with the observations \citep{Halekas05}. The $x$, $y$, $z$ axes are as defined in the aLSE co-ordinate system, where $x$-axis points into the plane of the paper.  }
		                         		   		    
		                         		   		   \label{fig:Ewake}
		                         \end{figure}

		To gain more insight into the entry process, we have carried out the backtracing of the protons belonging to different populations. The trajectory of the protons are computed backwards in time from their observed position in the lunar wake (in aLSE coordinates) under the Lorentz force. The position where the protons are observed in the lunar wake, and their observed velocities (in a given energy and direction bin of SWIM) are used as the initial values. Since the magnetic field in the wake has been found to increase to 1.5 $\times $ B$_{IMF}$ \citep{Colburn67}, for trajectory calculations, we used  magnetic field value of 1.5 $\times $ B$_{IMF}$  inside the wake region and 1.0 $\times $ B$_{IMF}$ outside the wake. Regarding the electric field, outside the wake, the protons experience the convective electric field of solar wind ($E_c =-V_{sw} \times B_{IMF}$). Inside the wake, protons will experience an additional electric field - the wake boundary electric field \citep{Halekas05,Kallio05,Nishino09a}. Similar to that discussed in \cite{Nishino09a} which agrees with the observations \citep{Halekas05}, we used a wake boundary electric field value of 0.7 mV m$^{-1}$ which acts in a region of width 0.25$\times$R$_L$  from the terminator, where $R_L$ refers to the lunar radius. Fig.~\ref{fig:Ewake} illustrates the geometry of the wake electric field used in the backtracing. As seen from Fig.~\ref{fig:Ewake}, the electric field is directed inward from the terminator towards the Sun-Moon line  \citep{Nishino09a}. 
		
			To solve the Lorentz equation numerically, Euler's method was used with a time step of 0.001 s. The model calculates the position and velocity of the protons back in time (in aLSE) until proton trajectory either intersect the lunar surface or travel a sufficiently larger distance ($>$ 5000 km) upstream. In other words, calculations terminate when either the magnitude of the position vector of the protons becomes less than or equal to the radius of the Moon (1738 km) or exceeds 5000 km. For the solar wind velocity and IMF, data from ACE satellite (time shifted to the location of Moon) was used. Backtracing was not done for the populations which were considered to be associated with Entry-1 (G2, H4, I1, L1).
		 
		\begin{figure}
		                     		   \centering
		                     		 		     \includegraphics[width=40pc]{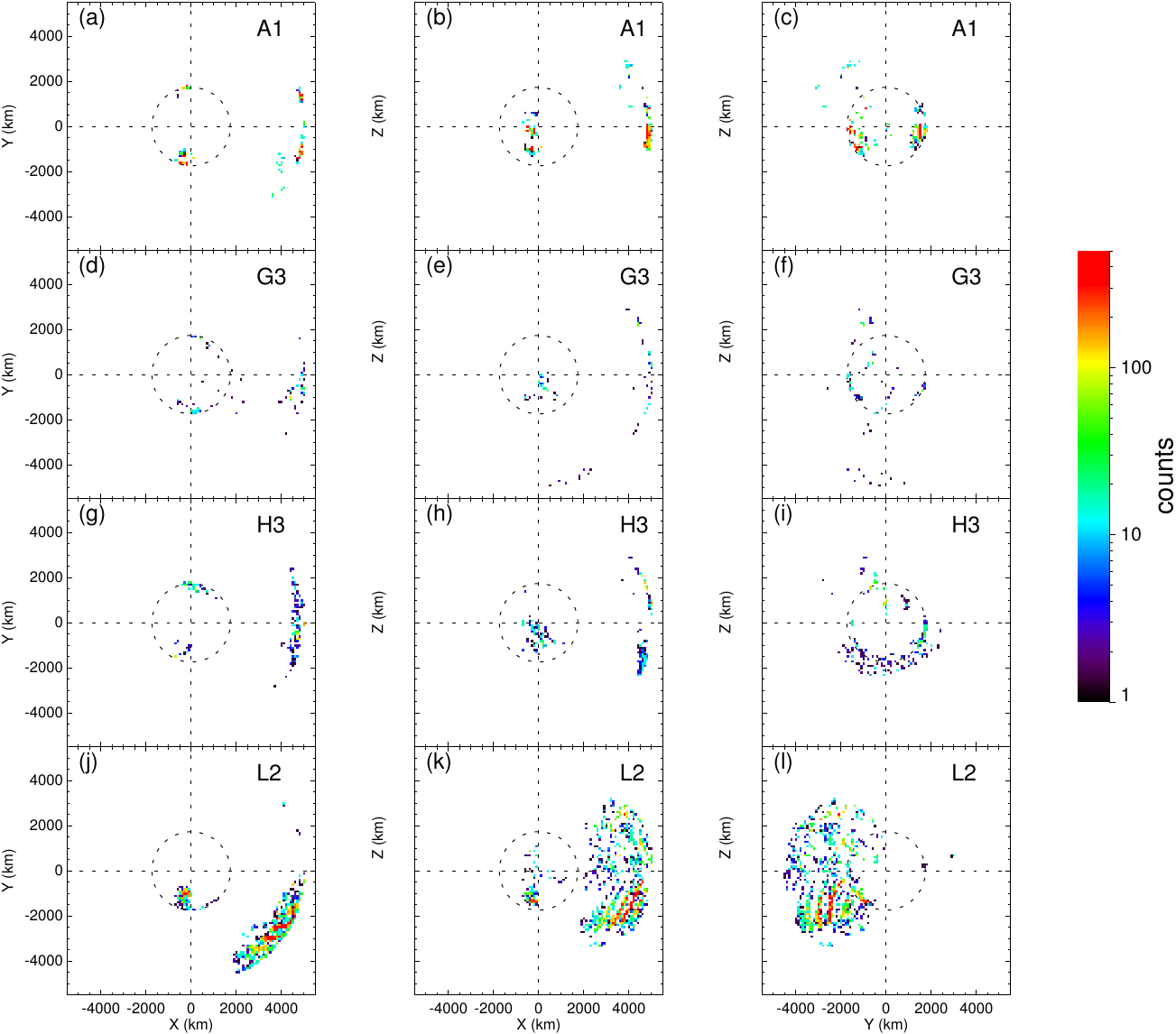}
		                     		   		    \caption{ Backtracing results in aLSE co-ordinates. (a)--(c): for population A1 in location 1, (d)--(f): population G3 in location 3, (g)--(i): population H3 in location 3, (j)--(l): population L2 in location 4.
		                     		   		    The left most panels (a), (d), (g), (j) show projection in the $x-y$ plane; middle panels (b), (e), (h), (k) show projection in the  $x-z$ plane; and right most panels (c), (f), (i), (l) represent projection in $y-z$ plane. The protons are traced back based on their observed location, energy and direction. The color bar represents the counts of the protons (observed in the specific energy and direction bins of SWIM) accumulated over the spatial grids of size 100 km $\times$ 100 km, at the end of backtracing. The protons which hit the nightside surface are unrealistic. The value of magnetic field in the wake is assumed to be 1.5$\times$ B$_{IMF}$ \citep{Colburn67}. In addition to the convective electric field of solar wind, a wake boundary electric field of 0.7 mV m$^{-1}$ has been considered (see Fig.~\ref{fig:Ewake} for details).  }
		                     		   		   \label{fig:backtraceCG}
		                     \end{figure}
		                     
		The results of the backtracing for the four populations A1, G3, H3, and L2 are shown in Fig.~\ref{fig:backtraceCG}. The plots are generated by binning the position co-ordinates of the protons at the end of backtracing into grids of size 100 $\times$ 100 km and are projected onto different 2-D planes ($x$-$y$, $x$-$z$, and $y$-$z$) in aLSE co-ordinates. The counts observed, in a particular energy and direction bin of SWIM, are accumulated over the spatial grids at the end of backtracing and is represented as color bar. The trajectories which terminate on the nightside surface should be ignored since they are unrealistic. 
		
		From Figs.~\ref{fig:backtraceCG}a to ~\ref{fig:backtraceCG}c, it can be seen that for population A1, almost all particles goes to the solar wind (99.3\%) with a minor contribution from the dayside surface (0.6\%) close to the terminator. The value 99.3\% represents the ratio of observed counts which terminate in solar wind to the sum of counts which terminate in the solar wind and on the dayside surface, at the end of backtracing. Similarly, 0.6\% represents the ratio of observed counts which terminate on the dayside surface to the sum of counts which terminate in solar wind and on the dayside surface, at the end of backtracing.  Thus, the source of A1 is most likely Entry-2 with minor contribution from Entry-5, as expected from the velocity distribution. 
		
		Figs.~\ref{fig:backtraceCG}d to ~\ref{fig:backtraceCG}f show that the population G3 has significant contribution (31\%) from the surface. The protons scattered from surface (or magnetic anomalies) may contribute significantly to G3. The mechanisms could be Entry-4 (from surface) and Entry-3 (from solar wind) as expected from the velocity distribution. Similarly, for the population H3 also (Figs.~\ref{fig:backtraceCG}g to \ref{fig:backtraceCG}i), backtracing shows significant contribution from the dayside surface (17\%). The mechanism could be Entry-4 which agrees with the inference drawn from the velocity distribution. Whereas both the velocity distribution and backtracing suggested the contribution from Entry-4, backtracing has shown that the dominant source is solar wind. The velocity distribution for H3 do not support Entry-2 as the source and Entry-3 is also not expected to contribute based on the prevailing IMF. Thus, the process by which solar wind contribute to H3 is left open. Population L2 (Figs.~\ref{fig:backtraceCG}j to \ref{fig:backtraceCG}l), have dominant contribution from solar wind with minor contribution (0.2\%) from surface. As discussed before, L2 was considered to be associated with Entry-5 and Entry-3 (partially) based on the velocity distribution. Backtracing shows that Entry-3 may still contribute whereas  Entry-5 may make only minor contribution to L2. 
		
		
		Similar backtracing carried out for other populations (results not shown), indicated that the population A2 has roughly equal  contributions from the solar wind (43.4\%) and the dayside surface (56.5\%). As discussed before, neither Entry-2 nor Entry-3 plays a role for A2. Thus the mechanism for the contribution from solar wind is not clear.  The surface contribution is more likely associated with Entry-5. For other populations in  location 1, backtracing showed that the source for B1, B2, B3, B4, C1, C2 and C3 is mainly solar wind with almost no contribution from surface.
		For B1, the velocity distribution indicate Entry-5 as possible mechanism whereas backtracing suggest no contribution from the surface.  The population B2 is more likely associated with Entry-2 and Entry-3, B3 with Entry-3, and C1 with Entry-2. As discussed before, Entry-3 may partially account for population C2, and hence there may be some other unknown entry mechanism that also contribute to C2.
		
		Regarding B4 and C3 in location 1, they have significant $v_\parallel$ and  $v_\perp$, and the velocities are comparable to that of solar wind. Also, the ($v_\parallel$, $v_\perp$) components are similar to that of solar wind. They were suspected to be associated with Entry-4 as inferred from the velocity distribution. However, backtracing suggested that they are purely of solar wind origin without any surface contribution (i.e., Entry-4 not supported). Thus their source remains unknown. However, since their velocity components are similar to that of solar wind, they could be lunar exospheric H$^+$ that are picked up by the solar wind (closer to the terminator) above the lunar surface. However, a detailed analysis is required to confirm these and is the scope for future work.
		
		In location 2, backtracing showed that populations D1 has minor contribution from surface ($<$1\%), and are dominantly of solar wind origin. They could be due to Entry-2 with minor contributions from Entry-5. Backtracing for D2 suggests that the the major source is solar wind with significant contribution from surface (21.5\%). The entry mechanism from solar wind could be Entry-3, whereas the mechanism for the source from the surface cannot be explained. Similar to D1, E1 also have dominant contribution from solar wind (99.5\%) with minor contribution from surface. The source of E1 could be Entry-2, Entry-3 and minor contribution from Entry-5 as expected. For E3, significant contribution is seen from surface ($\sim$7\%) and could be associated with Entry-3 as major process with significant role by Entry-5. F1 have solar wind as dominant source (98\%) and also contribution from surface. F1 could be due Entry-2, Entry-3 and with significant contribution from Entry-5.  
		  
		The population E2 in location 2, which appears as a strip in the velocity distribution map,  has comparable $v_\parallel$ and $ v_\perp$, and the  velocity magnitude is comparable to that of solar wind. The backtracing for E2 showed minor ($<$2\%) contribution from surface and dominant contribution from solar wind. For the surface contribution, Entry-4 across $-E_c$ pole is unlikely (even if the protons are from magnetic anomaly). Thus, the source for the surface contribution for E2 is not clear. For the solar wind part, Entry-3 may be the likely process. 
		
		In location 3, backtracing suggests that populations G1 and H1 have minor contribution from surface ($<$1\%), and  is dominantly of solar wind origin. Entry-2, Entry-3, and Entry-5 can be associated with these populations. For H2, backtracing showed major contribution from solar wind, likely to be due to Entry-3, and significant contribution from surface (11\%) which may be due to Entry-5. Backtracing for I2 suggests major contribution from solar wind  which may be due to Entry-3 with significant contribution from surface ($\sim$2.5\%), which could be due to Entry-4, as expected from the velocity distribution. 
		
		In location 4, for J1 backtracing shows solar wind as the source with no contribution from the surface. The entry mechanism for solar wind protons could be associated with Entry-3. For the population K1, the dominant source appears to be solar wind (likely to be associated with Entry-3) and the contribution from surface is negligible ($<$1\%) which can be associated with Entry-5.

		In summary, of the 28 populations of protons in the near-lunar wake observed by SWIM, several populations are associated with known processes, such as Entry-1, Entry-2, Entry-3, Entry-4 and Entry-5.  There are a few populations whose entry process could not be explained by the known mechanisms. These include direct solar wind contribution for A2, B1, B4, C3, E2, H3, and dayside surface contribution for D2. Since Entry-3 may partially account for populations C2 and L2, there may be some other unknown entry mechanism that also contribute to these populations. Table~\ref{table:summaryVDFBT} provides a summary of the results on the entry mechanisms associated with the different populations that are arrived at based on the velocity distribution and backtracking.
		  
		   It is to be noted that except parallel entry, all other  known entry mechanism involve the gyration of the solar wind protons around IMF, either by virtue of large thermal velocity or by experiencing additional electric field (convective or at wake boundary) and their trajectories enable the protons to enter the near-lunar wake.
		   
		   \begin{sidewaystable}
		   \caption{Entry mechanisms expected from velocity distribution, the source identified from backtracing, and inferences on the entry mechanisms of the different populations at locations 1--4 compared with that of backtracing}
		   \label{table:summaryVDFBT}
		   \centering
		   \begin{tabular}{lllll}
		   \hline
		   Location & population  & Entry mechanism from &  Result from backtracing* & Inference  \\
		   & &  velocity distribution&  &\\
		    \hline
		    1 & A1 &  Entry-2, Entry-5 & solar wind (99.3\%)  & Entry-2, Entry-5\\
		    1 & A2 &  Entry-5 & solar wind (43.4\%)  & Entry-5 and unknown mechanism\\
		    & & & &  for the direct solar wind entry \\
		    1 & B1 &  Entry-5 & solar wind (100\%) & mechanism unknown\\
		     1 & B2 &  Entry-2, Entry-3, Entry-5 & solar wind (100\%) & Entry-2,Entry-3\\
		      1 & B3 & Entry-3, Entry-5 & solar wind (100\%) & Entry-3\\
		      1 & B4 & Entry-4 &  solar wind (100\%) & unknown\\
		      1 & C1 & Entry-2, Entry-5 &  solar wind (100\%)  & Entry-2\\   
		    1 & C2 & Entry-3** &  solar wind (100\%) & Entry-3 and unknown mechanism\\
		    1 & C3 & Entry-4 &  solar wind (100\%) & unknown \\    
		    2 & D1 & Entry-2, Entry-5 &  solar wind (96.7\%)  & Entry-2, Entry-5\\   
		    2 & D2 & Entry-3 &  solar wind (78.5\%)  & Entry-3, and source from the  \\
		    & & & & dayside surface is unknown\\ 
		    2 & E1 & Entry-2, Entry-3, Entry-5 &  solar wind (99.5\%)  & Entry-2, Entry-3, Entry-5\\   
		    2 & E2 & unknown  &  solar wind (98.3\%)  &  process unknown\\
		    2 & E3 & Entry-3, Entry-5  &  solar wind (93\%)  & Entry-3, Entry-5\\
		    2 & F1 & Entry-2, Entry-3, Entry-5  &  solar wind (98\%)  & Entry-2, Entry-3, Entry-5\\
		    3 & G1 & Entry-2, Entry-3, Entry-5  &  solar wind (99.6\%)  & Entry-2, Entry-3, Entry-5\\
		    3 & G2 & Entry-1 &  --- & Entry-1\\
		    3 & G3 & Entry-3, Entry-4  &  solar wind (69\%)  & Entry-3, Entry-4\\
		    3 & H1 & Entry-2, Entry-3, Entry-5  &  solar wind (98.6\%) & Entry-2, Entry-3, Entry-5\\
		    3 & H2 & Entry-3, Entry-5  &  solar wind (89\%)  & Entry-3, Entry-5\\
		    3 & H3 & Entry-4  &  solar wind (83\%)  & Entry-4 and unknown mechanism\\
		     & & & &  for the direct solar wind entry \\
		    3 & H4 & Entry-1  & --- & Entry-1\\
		    3 & I1 & Entry-1  & ---  & Entry-1\\
		    3 & I2 & Entry-3, Entry-4  &  solar wind (97.5\%) & Entry-3, Entry-4\\
		    4 & J1 & Entry-3, Entry-5  &  solar wind (100\%) & Entry-3\\
		   4 & K1 & Entry-3, Entry-5  &  solar wind (99\%) & Entry-3, Entry-5\\
		   4 & L1 & Entry-1  &  solar wind (99.9\%)  & Entry-1\\
		   4 & L2 & Entry-3**, Entry-5  &  solar wind (99.8\%) & Entry-3 and unknown mechanism, Entry-5\\

		   %
		    \hline
		   \end{tabular}
		   \flushleft {\footnotesize * The contribution from solar wind is given. The value in brackets represents the ratio (in percentage) of the observed proton counts which terminate in solar wind to the sum of the counts which terminate in the solar wind and the dayside surface, at the end of backtracing. The difference of this value from 100\% represents the contribution from the dayside surface.\\
		   ** Partially due to the low IMF condition during some of the observation of these populations }
		   \end{sidewaystable}

\section{Conclusion}

Using the observations by SWIM sensor of the SARA experiment on Chandrayaan-1, the velocity distributions of the protons in the near-lunar wake (100 to 200 km from the lunar surface) are computed in the solar wind rest frame.  The velocity distributions were further sorted according to the angle between the upstream solar wind velocity and IMF. The distributions are not identical for different IMF orientation, which suggests the control of IMF in the proton entry process. The protons were found to enter the lunar wake parallel as well as perpendicular to the direction of IMF.  Most of the observed protons have velocities lower than that of the solar wind, and a few populations have velocity comparable to that of solar wind. From the velocity distribution, several population of protons were identified based on the selection criterium. The possible mechanism for the proton entry to the near wake associated with these populations were inferred from the characteristics of velocity distribution. The mechanisms include diffusion of solar wind protons into the wake along IMF, by virtue of gyro-radii of solar wind protons and scattering of solar wind protons from the dayside lunar surface or from magnetic anomalies. To gain more insight into the source of these populations,  the trajectory of the protons are computed back in time (backtracing) under the influence of IMF, convective and wake boundary electric field. For most of the populations, the source of the protons  obtained from backtracing agree with the inferences drawn from the velocity distribution. There are few populations whose entry mechanism could not be explained by the known processes.



\flushleft \textbf {Acknowledgments}\\[5pt]
\justify {We thank the ACE SWEPAM instrument team, ACE MAG instrument team and the ACE Science Center for providing the ACE data. This work has been supported by the Indo-Swedish International Collaborative Research Grant by the  Swedish International Developement Corporation Agency (SIDA). The efforts at Space Physics Laboratory of Vikram Sarabhai Space Centre are supported by Indian Space Research Organisation (ISRO). The effort at the University of Bern was supported in part by ESA and by the Swiss National
Science Foundation.}




\flushleft \textbf {References}

\clearpage

 
\clearpage


\end{document}